\documentstyle[sprocl]{article}
\begin{document}\bibliographystyle{unsrt}\arraycolsep1.5pt\begin{titlepage}
\renewcommand{\thefootnote}{\fnsymbol{footnote}}\hfill\begin{tabular}{l}
HEPHY-PUB 733/00\\UWThPh-2000-38\\hep-ph/0010078\\September 2000\end{tabular}
\\[1cm]\large\begin{center}{\bf INSTANTANEOUS BETHE--SALPETER EQUATION:
(SEMI-)ANALYTICAL SOLUTION}\\\vspace{0.7cm}\large{\bf Wolfgang
LUCHA\footnote[1]{\normalsize\ {\em E-mail address\/}:
wolfgang.lucha@oeaw.ac.at}}\\[.2cm]\normalsize Institut f\"ur
Hochenergiephysik,\\\"Osterreichische Akademie der
Wissenschaften,\\Nikolsdorfergasse 18, A-1050 Wien,
Austria\\[0.5cm]\large{\bf Khin MAUNG MAUNG\footnote[3]{\normalsize\ {\em
E-mail address\/}: maung@jlab.org}}\\[.2cm]\normalsize Department of Physics,
Hampton University,\\Hampton, VA 23668\\[0.5cm]\large{\bf Franz F.
SCH\"OBERL\footnote[2]{\normalsize\ {\em E-mail address\/}:
franz.schoeberl@univie.ac.at}}\\[.2cm]\normalsize Institut f\"ur Theoretische
Physik, Universit\"at Wien,\\Boltzmanngasse 5, A-1090 Wien,
Austria\\[.6cm]\abstracts{\em Presented by F.~Sch\"oberl at the International
Conference ``Quark Confinement~and the Hadron Spectrum IV,'' July 3 -- 8,
2000, Vienna, Austria\/}\normalsize{\bf
Abstract}\end{center}\vspace{3ex}\abstracts{The Bethe--Salpeter equation for
bound states of a fermion--antifermion pair in the instantaneous
approximation for the involved interaction kernel is converted~into~an
equivalent matrix eigenvalue problem with explicitly (algebraically) given
matrices.}\normalsize\noindent{\em PACS numbers\/}: 11.10.St, 03.65.Ge
\renewcommand{\thefootnote}{\arabic{footnote}}\end{titlepage}

\title{INSTANTANEOUS BETHE--SALPETER EQUATION: (SEMI-)ANALYTICAL
SOLUTION}\author{Wolfgang LUCHA}\address{Institute for High Energy Physics,
Austrian Academy of Sciences,\\Nikolsdorfergasse 18, A-1050 Vienna,
Austria\\E-mail: wolfgang.lucha@oeaw.ac.at}\author{Khin MAUNG
MAUNG}\address{Department of Physics, Hampton University,\\Hampton, VA
23668\\E-mail: maung@jlab.org}\author{Franz F.~SCH\"OBERL}\address{Institute
for Theoretical Physics, University of Vienna,\\Boltzmanngasse 5, A-1090
Vienna, Austria\\E-mail:
franz.schoeberl@univie.ac.at}\maketitle\abstracts{The Bethe--Salpeter
equation for bound states of a fermion--antifermion pair in the instantaneous
approximation for the involved interaction kernel is converted~into~an
equivalent matrix eigenvalue problem with explicitly (algebraically) given
matrices.}

\section{The Instantaneous Bethe--Salpeter Equation (IBSE)}For a system of
massless fermion and antifermion forming bound states with~the ``pion-like''
spin, parity, and charge conjugation quantum numbers $J^{PC}=0^{-+},$ the
(homogeneous) Bethe--Salpeter equation, in free-propagator approximation and
instantaneous approximation for the involved interaction kernel, reads for~a
time-component Lorentz vector interaction (i.e., the Dirac structure
$\gamma^0\otimes\gamma^0$)~\cite{Lagae,Olsson}\begin{eqnarray}
2\,k\,\Psi_2(k)+\int\limits_0^\infty\frac{{\rm d}k'\,k'^2}{(2\pi)^2}\,
V_0(k,k')\,\Psi_2(k')&=&M\,\Psi_1(k)\ ,\nonumber\\2\,k\,\Psi_1(k)
+\int\limits_0^\infty\frac{{\rm d}k'\,k'^2}{(2\pi)^2}\,V_1(k,k')\,\Psi_1(k')
&=&M\,\Psi_2(k)\ .\label{Eq:IBSEm0}\end{eqnarray}In this set of coupled
equations for the two relevant radial Salpeter amplitudes $\Psi_1$ and
$\Psi_2$ in momentum space, with the bound-state masses $M$ as eigenvalues,
the interaction potential $V(r),$ usually formulated in configuration space,
enters in form of its standard Fourier--Bessel transforms $V_L(k,k'),$
$L=0,1.$ We adopt a linear potential $V(r)=\lambda\,r$ ($\lambda>0)$ as a
simple model for quark confinement.

\section{Efficient Method of Solution: Expansion in Terms of Basis States}By
insertion of the first of Eqs.~(\ref{Eq:IBSEm0}) into the second and by
expansion in terms~of sets (distinguished by the angular momenta $\ell=0,1$)
of basis states for~$L_2(R^+)$ ---with configuration and momentum-space
representations $\phi_i^{(\ell)}(r)$ and $\widetilde\phi_i^{(\ell)}(p),$
resp.---the solution of the IBSE~(\ref{Eq:IBSEm0}) reduces to the
diagonalization of the~matrix\cite{Lucha00:IBSEm0}\begin{eqnarray}{\cal
M}_{ij}&=&4\int\limits_0^\infty{\rm d}k\,k^4\,
\widetilde\phi_i^{(0)}(k)\,\widetilde\phi_j^{(0)}(k)
+2\int\limits_0^\infty{\rm d}k\,k^3\,\widetilde\phi_i^{(0)}(k)
\int\limits_0^\infty\frac{{\rm d}k'\,k'^2}{(2\pi)^2}\,V_0(k,k')\,
\widetilde\phi_j^{(0)}(k')\nonumber\\ &+&2\int\limits_0^\infty{\rm
d}k\,k^2\,\widetilde\phi_i^{(0)}(k)
\int\limits_0^\infty\frac{{\rm d}k'\,k'^3}{(2\pi)^2}\,V_1(k,k')\,
\widetilde\phi_j^{(0)}(k')\nonumber\\ &+&\int\limits_0^\infty{\rm
d}k\,k^2\,\widetilde\phi_i^{(0)}(k)
\int\limits_0^\infty\frac{{\rm d}k'\,k'^2}{(2\pi)^2}\,V_1(k,k')
\int\limits_0^\infty\frac{{\rm d}k''\,k''^2}{(2\pi)^2}\,V_0(k',k'')\,
\widetilde\phi_j^{(0)}(k'')\ .\label{Eq:M2}\end{eqnarray}Allowing these basis
functions to depend on a variational parameter $\mu>0$~gives us more freedom
in the search for solutions of the IBSE. All integrations~in~${\cal M}_{ij}$
are evaluated by (truncated) expansions, with the ($\mu$-independent)
coefficients~\cite{Lucha00:IBSEm0}\begin{eqnarray*}
&&I^{(2)}_{ij}\equiv\frac{1}{\mu^2}\int\limits_0^\infty{\rm
d}k\,k^4\,\widetilde\phi_i^{(0)}(k)\,\widetilde\phi_j^{(0)}(k)\ ,\quad
i,j=0,1,2,\dots\ ,\\ &&b_{ij}\equiv\frac{1}{\mu}\int\limits_0^\infty{\rm
d}k\,k^3\,\widetilde\phi_i^{(0)}(k)\,\widetilde\phi_j^{(0)}(k)\ ,\quad
k\,\widetilde\phi_i^{(0)}(k)
=\mu\,\sum_{j=0}^N\,b_{ji}\,\widetilde\phi_j^{(0)}(k)\ ,\\
&&c_{ij}\equiv\int\limits_0^\infty{\rm
d}k\,k^2\,\widetilde\phi_i^{\ast(1)}(k)\,\widetilde\phi_j^{(0)}(k)\ ,\quad
\widetilde\phi_i^{(0)}(k)=\sum_{j=0}^N\,c_{ji}\,\widetilde\phi_j^{(1)}(k)\
,\\ &&d_{ij}\equiv\frac{1}{\mu}\int\limits_0^\infty{\rm d}k\,k^3\,
\widetilde\phi_i^{\ast(1)}(k)\,\widetilde\phi_j^{(0)}(k)\ ,\quad
k\,\widetilde\phi_i^{(0)}(k)
=\mu\,\sum_{j=0}^N\,d_{ji}\,\widetilde\phi_j^{(1)}(k)\ ,\\
&&V^{(\ell)}_{ij}\equiv\mu\int\limits_0^\infty{\rm
d}r\,r^3\,\phi_i^{(\ell)}(r)\,\phi_j^{(\ell)}(r)\ ,\quad
r\,\phi_i^{(\ell)}(r)=\frac{1}{\mu}\,\sum_{j=0}^N\,V^{(\ell)}_{ji}\,
\phi_j^{(\ell)}(r)\ ,\quad\ell=0,1\ .\end{eqnarray*}The explicit algebraic
expressions of all these matrices may be found in Ref.~3.\footnote{Let's
mention a numerical problem noted for Mathematica 4.0: for, e.g., the
matrix~element $d_{49,49}=101/(2\sqrt{689})=1.924$ Mathematica finds exactly
this value for a working precision~of 40 digits but the nonsense value
$-1.675\times 10^{22}$ for the default working precision of 16 digits.}

\newpage\noindent In this way, the IBSE~(\ref{Eq:IBSEm0}) is converted to an
eigenvalue problem for the matrix~\cite{Lucha00:IBSEm0}\begin{eqnarray*}{\cal
M}_{ij}&=&4\,\mu^2\,I^{(2)}_{ij}+2\,\lambda\,\sum_{r=0}^N\,b_{ri}\,V^{(0)}_{rj}
+2\,\lambda\,\sum_{r=0}^N\,\sum_{s=0}^N\,c^\ast_{ri}\,d_{sj}\,V^{(1)}_{rs}\\
&+&\frac{\lambda^2}{\mu^2}\,\sum_{r=0}^N\,\sum_{s=0}^N\,\sum_{t=0}^N\,
c^\ast_{ri}\,c_{st}\,V^{(1)}_{sr}\,V^{(0)}_{tj}\ .\end{eqnarray*}

\section{Analytical Results (for Both Massless and Massive Constituents)}For
a matrix size less than or equal to 4, the diagonalization of the
matrix~${\cal M}_{ij}$ may be even performed analytically. In the
one-dimensional case, we find, after minimizing w.r.t.\ the variational
parameter $\mu,$ for the lowest bound-state~mass~\cite{Lucha00:IBSEm0}$$
M=4\,\sqrt{\frac{2\,\lambda}{3\pi}\left(2+\sqrt{5}\right)}\ .$$
For~\cite{Lucha91} $\lambda=0.2\;\mbox{GeV}^2,$ this expression gives
$M=1.696\;\mbox{GeV},$ only 2.4\,\% away~from the numerical result
$M=1.656\;\mbox{GeV},$ obtained for $15\times 15$ matrices and $N=49.$ For a
nonvanishing mass $m$ of the bound-state constituents, we get
accordingly$$M^2=8\,m^2+\frac{8896}{315\,\pi}\,\lambda
+\frac{23}{7}\left(\frac{128\,\lambda}{45\,\pi\,m}\right)^2\qquad(m\ne0)\ .$$

\section{Relations Between Matrix Elements and Accuracy of Expansions}Our
final question concerns the errors induced by the necessary truncations~of
the expansion series. The expansion coefficients $b_{ij},$ $c_{ij},$ $d_{ij}$
are not independent but should satisfy (clearly, only in the limit
$N\to\infty$ exact) relations of
the~kind$$\sum_{r=0}^N\,c^\ast_{ri}\,c_{rj}=\delta_{ij}\
,\quad\sum_{r=0}^N\,c^\ast_{ri}\,d_{rj}=\sum_{r=0}^N\,d^\ast_{ri}\,c_{rj}=
b_{ij}\ ,\quad\sum_{r=0}^N\,d^\ast_{ri}\,d_{rj}=I^{(2)}_{ij}\ .$$For
$15\times 15$ matrices and $N=49,$ these relations are fulfilled with
relative~errors less than~$3\,\%.$ For comparison, some integrals in
(\ref{Eq:M2}) may be evaluated exactly.~\cite{Lucha00:IBSEm0}

\end{document}